\documentclass[%
reprint,
superscriptaddress,
preprintnumbers,
nofootinbib,
amsmath,amssymb,
aps,
]{revtex4-1}

\usepackage{graphicx}
\usepackage{dcolumn}

\usepackage{changes}
\usepackage{slashed}

\def\be{\begin{equation}}
\def\ee{\end{equation}}

\def\l{\left(}
\def\r{\right)}

\renewcommand{\ln}{\mathop{\rm ln}\nolimits}

\newcommand{\Tr}{{\rm Tr}}
\newcommand{\bg}{\begin{gather}}
\newcommand{\eg}{\end{gather}}

\begin{document}

\preprint{INR-TH-2016-048}

\title{Prospects of models with light sgoldstino in electron beam dump experiment at CERN SPS}

\author{K.\,O.\,Astapov}
\email{astapov@ms2.inr.ac.ru}
\affiliation{Institute for Nuclear Research of the Russian Academy of Sciences,
  Moscow 117312, Russia}%
\affiliation{Physics Department, Moscow State University, Vorobievy Gory,  
Moscow 119991, Russia}

\author{D.\,V.\,Kirpichnikov}
\email{kirpich@ms2.inr.ac.ru}
\affiliation{Institute for Nuclear Research of the Russian Academy of Sciences,
  Moscow 117312, Russia}%

\begin{abstract}
We discuss phenomenology of light scalar sgoldstino in context of CERN electron beam dump experiment 
NA64. We calculate sgoldstino production rate for this experiment taking into account sgoldstino mixing with the Higgs 
boson and find a region in the model parameter space which can be tested in NA64. 
\end{abstract} 

\maketitle

\section{Motivation}
\label{sec:Intro}
Supersymmetry is a promising  extention  of the Standard Model~\cite{Haber:1984rc,Martin:1997ns}.
However according to experimentally observed absence of superpartners at low energies, SUSY models imply supersymmetry to be spontaneously broken at some scale. The breaking mechanism is provided  by  underlying
microscopic theory.
The breaking  can happen when a hidden sector dynamics  results in  a nonzero
vacuum expectation value $F$ to an auxiliary component
of the superfield \cite{Brignole:2003cm}  and $\sqrt{F}$ is  SUSY breaking scale.
According to supersymmetric analog of the Goldstone theorem
\cite{Volkov:1972jx} there should exist a massless fermionic degree of freedom, goldstino.

In the simplest case this fermion belongs to a chiral multiplet, dubbed goldstino supermultiplet.
Apart from goldstino, it contains scalar - sgoldstino and an auxiliary field acquiring nonzero vacuum expectation value $F$, which trigger spontaneous SUSY breaking. 
The quantity $\sqrt{F}$ is supersymmetry breaking scale. Couplings of sgoldstino to SM fields are 
suppressed by $\frac{1}{F}$ and expected to be quite small. 
Being included into supergravity framework
goldstino becomes longitudinal component of gravitino with
mass related to the scale of supersymmetry breaking $\sqrt{F}$ as follows $m_{3/2} = \frac{F}{\sqrt{3}M_{pl}},$ where $M_{pl}$ is the Planck mass \cite{Cremmer:1978iv}. In the present work we consider the mass of sgoldstino to be light (less than $1$ GeV) as phenomenologically interesting case;  therefore a high intensity beam is
required to test the model via production of  sgoldstino. 
Beam dump experiment can perform the task. A preliminary estimate of the search for  sgoldstino  at beam dump experiment SHiP for sgoldstino production  and decay mechanisms can be found
in Ref.\,\cite{Astapov:2015otc}; here we complete that study by considering production of sgoldstino in electron-proton collisions.

The purpose of the present paper is to estimate the signal rate of sgoldstino decays expected to detect at
the NA64 electron beam-dump experiment in context of present model (see Refs.\,\cite{Andreas:2013lya,Banerjee:2016tad, Gninenko:2016kpg} for description of the experiment in context of the search for Dark photon). 
We consider the case when mostly the sgoldstino decays into
electron-positron pairs as  searching signature.

The paper is organized as follows.  In section \ref{sec:Lagrangian} we present effective interaction Lagranginan of sgoldstino to SM particles and set of chosen values of parameters of MSSM. Section \,\ref{sec:3} contains calculation of sgoldstino production cross section. In section \,\ref{sec:4} we discuss sgoldstino decay channels and calculate its lifetime for given model parameters.  
In Secs. \,\ref{sec:5} we calculate NA64 sensitivity to the SUSY breaking scale and put new limits on the model parameters. We conclude in Sec.\,\ref{conclude} by summarizing the results obtained.

\section{Lagrangian and parameters}
\label{sec:Lagrangian}
  
To the leading order in $1/F$, sgoldstino 
couplings to SM gauge fields - 
photons $F_{\mu\nu}$, gluons $G_{\mu\nu}$ and matter fields - leptons $l_a$, up and
down quarks $u_a$ and $d_a$, where index $a$ runs over three gererations at the mass scale above $\Lambda_{QCD}$ but below electroweak symmetry breaking reads as \cite{Astapov:2014mea, Gorbunov:2000th}

\begin{widetext}
\begin{multline}
{\cal L}^{s}_{eff}=-\frac{M_{\gamma\gamma}}{2\sqrt{2}F}sF^{\mu\nu}F_{\mu\nu}-\frac{M_2}{\sqrt{2}F}sW^{\mu\nu}W_{\mu\nu}-\frac{M_{ZZ}}{2\sqrt{2}F}sZ^{\mu\nu}Z_{\mu\nu}- \frac{M_3}{2\sqrt{2}F}s{}\Tr{}G^{\mu\nu}G_{\mu\nu}-\frac{A^{U}_{ab}v}{\sqrt{2}F}su_{a}u_{b}-\frac{A^{D}_{ab}v}{\sqrt{2}F}sd_{a}d_{b}-\\
-\frac{A_{ab}^{L}v}{\sqrt{2}F}sl_{a}l_{b}
\label{eef}
\end{multline}
\end{widetext}
Here $M_3$ is the gluino mass,
$M_{\gamma\gamma}=M_1\sin^2\theta_W+M_2\cos^2\theta_W$ and $M_{ZZ}=M_1\cos^2\theta_W+M_2\sin^2\theta_W$ with $M_1$ and
$M_2$ being $U(1)_Y$- and $SU(2)_W$-gaugino masses and $\theta_W$
the weak mixing angle, and  
$A^{U}_{ab}$, $A^{D}_{ab}$ and $A_{ab}^{L}$ are soft trilinear coupling constants. Lagrangian \eqref{eef}
includes only single-sgoldstino  interaction terms; considered in 
Refs.\,\cite{Perazzi:2000id,Perazzi:2000ty,Gorbunov:2000th,Demidov:2011rd},  double-sgoldstino terms are
suppressed by $1/F^2$ and are not probable for testing at the NA64
experiment. 


In general sgoldstino also mixes with neutral Higgs bosons as discussed in
Refs.\,\cite{Dudas:2012fa,Bellazzini:2012mh,Astapov:2014mea,Sobolev:2016gmr}: the 
scalar sgoldstino $S$ mixes with neutral
light $h$ and heavy $H$ Higgs bosons, while pseudoscalar $P$ mixes
with their axial partner $A$. We account for the 
mixing with $h$ only, since the other two do not change light scalar sgoldstino phenomenology
at NA64 for  considered set of parameters of the model.
Mixing of  the scalar sgoldstino
and  the lightest MSSM Higgs boson (SM-like Higgs) $h$ can be written 
as\,\cite{Astapov:2014mea}
\begin{equation}
\label{L-mixing}
{\cal L}_{mixing}=\frac{X}{F}\,Sh\,,
\end{equation}
where the mixing parameter $X$ is
\begin{equation}
\label{X_mixing}
X = 2\mu^3v\sin{2\beta} +
\frac{1}{2}v^3(g_1^2M_1+g_2^2M_2)\cos^2{2\beta}\,, 
\end{equation}
here $\mu$ is Higgsino mixing mass
parameter, $v=174$\,GeV is the Higgs vacuum expectation value (vev), 
$\tan\beta$ is describing the Higgs vev ratio, and $g_2$ and $g_1$ are 
$SU(2)_W$ and $U(1)_Y$ gauge coupling constants.

Since we are  considering sgoldstino $S$ mass  to be less than $1$ GeV (much lighter than the SM-like
Higgs boson of mass $m_h\approx125$\,GeV) all the Higgs-like couplings of scalar resonanse are  
suppressed by the mixing angle 
 \begin{equation}
\label{mix_angle}
\theta =-\frac{X}{Fm_{h}^2}\,.
\end{equation}

In Table\,\ref{MSSMpoint} we set numerical values for parameters of the MSSM  so
that $h$ asquire its experimentally observed value of $125$\,GeV by loop corrections from squark masses and trilinear couplings; and $H$ along with $A$ fields acquire heavy masses over $1$\,TeV to 
not to contribute into mixing with scalar and pseudoscalar sgoldstinos. In this arbitrary choice we suppose that all the model parameters 
take experimentally allowed values.

\begin{table}[!htb]
\begin{center}
\begin{tabular}{|l|l|l|l|l|}
\hline
$M_1,$ GeV &$M_2,$ GeV&$M_3,$ GeV&$\mu,$ GeV&$\tan\beta$\\ \hline
100&250&1500&1000&6\\ \hline \hline
$m_A,$ GeV&$A_l,$ GeV&$m_l,$ GeV&$A_Q,$ GeV&$m_Q,$ GeV\\ \hline
1000&2800&1000&2800&1000\\ \hline
\end{tabular}
\caption{\label{MSSMpoint}MSSM benchmark point.}
\end{center}
\end{table}
In the table we denoted $A^U_{aa}$, $A^D_{aa}$ as $A_Q$ and $A^L_{aa}$ as $A_l$,
all the off-diagonal $A^{U,D,L}_{ab}$ are set to zero.

\section{Production mechanism}
\label{sec:3}
\begin{figure}[htb!]
   \centering
\includegraphics[width=0.5\textwidth]{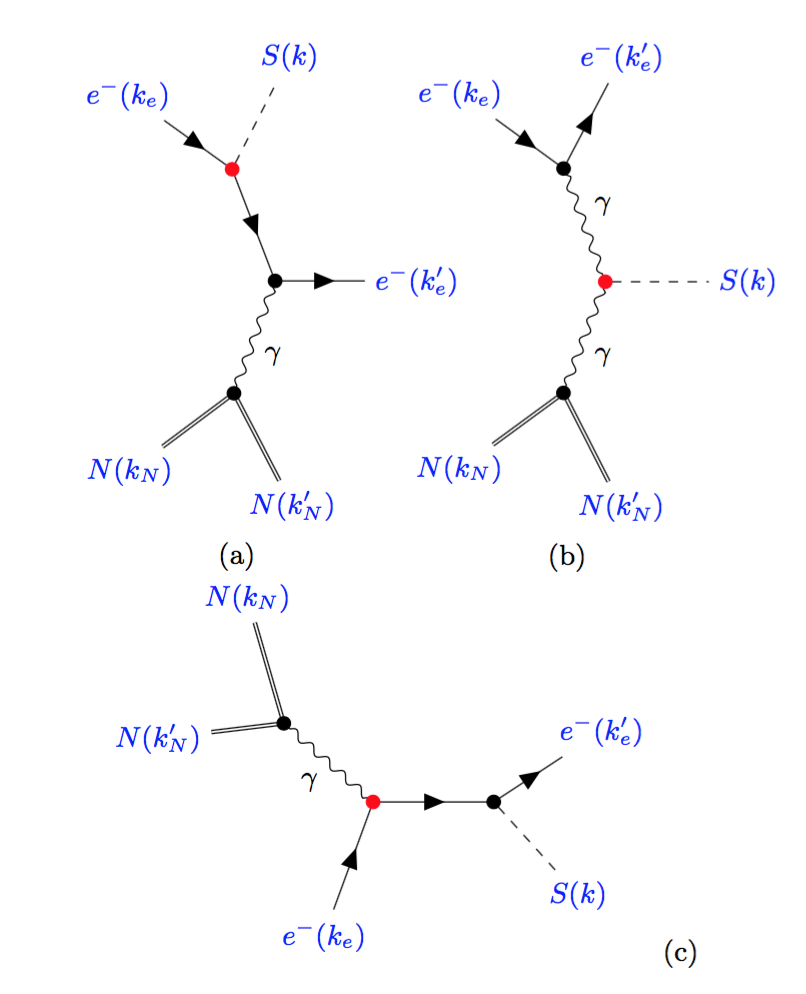}
\caption{Feynman diagrams describing $2\to3$ sgoldstino  production process.
\label{feynD}}  
\end{figure}

In this section we describe scalar sgoldstino production in electron-proton collisions as $100$ GeV electron beam hitting heavy nuclei lead ($Z = 82$) target. We take into account interactions of sgoldstino with nuclei and electrons. Feynman diagrams of the process are presented on Fig.\,\ref{feynD}. We denote the four-momenta of the initial beam and scattered electrons by $k_e=(E_e, \vec{k_e})$
and $k_e^{\prime} = (E_e^{\prime}, \vec{k_e^{\prime}})$; the four-momenta of the initial and final target state by $k_N^{\prime} = (E_N^{\prime}, \vec{k_N})$ and $k_N^{\prime} = (E_N^{\prime}, \vec{k_N^{\prime}})$; for outgoing sgoldstino particle $k=(E, \vec{k})$.
Expressions for corresponding diagrams read as:

\begin{multline}
i\mathcal{M}^{(b)}=\frac{M_{\gamma\gamma}}{2\sqrt{2}F}(-ie)^2\frac{i}{(k_p-k_p^{\prime})^2}\frac{i}{(k_e-k_e^{\prime})^2}\times\\
\times\Big[-2(k_e-k_e^{\prime},k_p-k_p^{\prime})g^{\alpha\beta}+\\
(k_e-k_e^{\prime})^{\alpha}(k_p-k_p^{\prime})^{\beta}+(k_e-k_e^{\prime})^{\beta}(k_p-k_p^{\prime})^{\alpha}\Big]j^e_{\alpha}j^N_\beta,
\end{multline}

\begin{equation}
i\mathcal{M}^{(a)}+i\mathcal{M}^{(c)}=\frac{A_Lv}{\sqrt{2}F}\times(-ie)^2\frac{i}{(k_p-k_p^{\prime})^2}L_{\alpha}j^N_{\alpha},
\end{equation}

where leptonic and hadronic  currents read as  
$$j^e_{\alpha} = \bar{u}_e(k_e)\gamma_{\alpha}u(k_e)$$
and 

$$j^N_{\beta} = Z F(Q_t)(k_p+k_p^{\prime})_{\alpha}$$
correspondingly. 
Here $F(Q)$ is the nuclear charge form factor \cite{Beranek:2013nqa}. Note, that we did not
consider diagrams similar to (a), (b) and (c) but with $Z^0$-boson exchange, since they are suppressed
by $Z^0$ mass.

Leptonic tensor:
$$L_{\alpha}=\bar{u}_e(k_e)\Bigg(\gamma_{\alpha}\frac{-(\slashed{k}_e-\slashed{k})+m_e}{(k_e-k)^2-m_e^2}+\frac{-(\slashed{k}+\slashed{k}_e^{\prime})+m_e}{(k+k_e^{\prime})^2-m_e^2}\gamma_{\alpha}\Bigg)u(k_e),$$

Therefore full amplitude of the process reads as
\begin{equation}
i\mathcal{M}=i\mathcal{M}^a+i\mathcal{M}^b+i\mathcal{M}^c
\end{equation}

\begin{figure}[htb!]
   \centering
\includegraphics[width=0.5\textwidth]{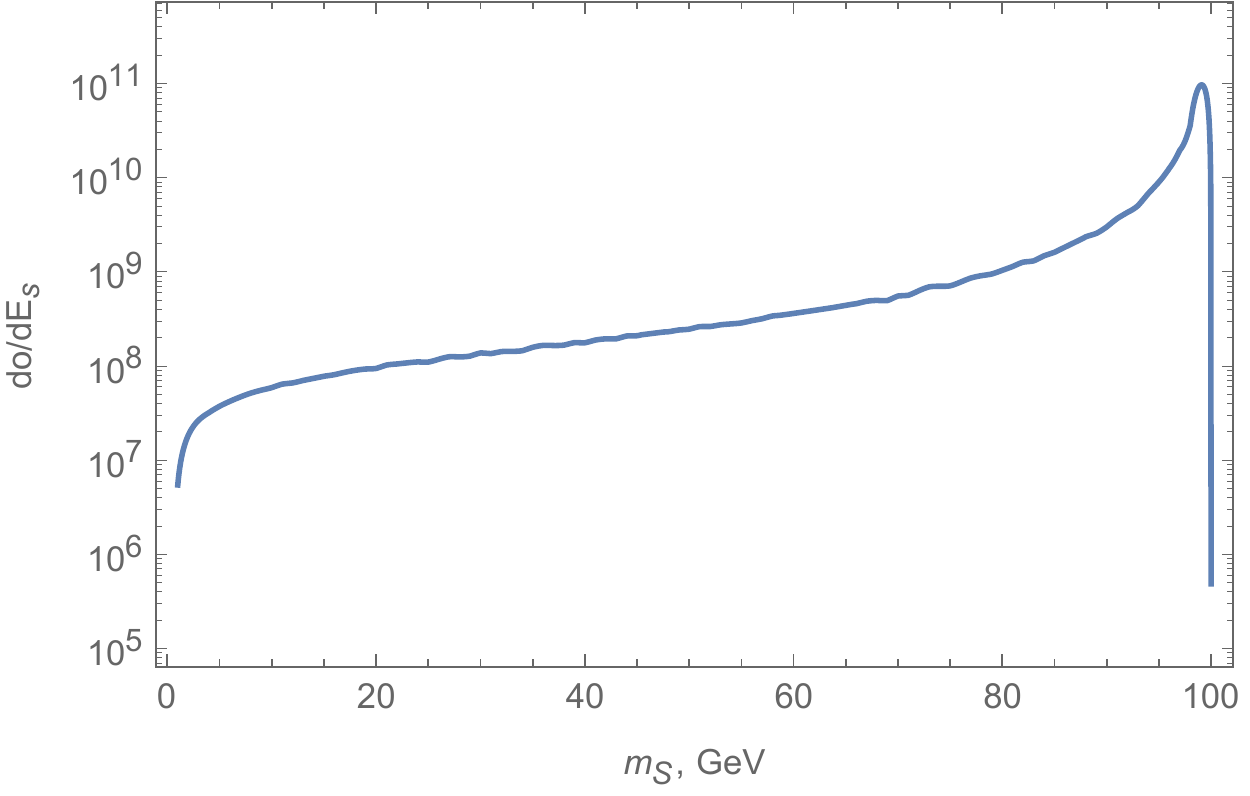}
\caption{Differential production cross section of sgoldstino as a function of its mass. Mass 
of the sgoldstino in taken 100 MeV.
\label{CS}}  
\end{figure}

The differential cross section of $2\to3$ process for $m_S=100$ MeV is presented on Fig.\ref{CS} 

\section{Decay channels}
\label{sec:4}

For (sub-)GeV mass-range sgoldstino decay channels into 
pairs of SM particles, if kinematically allowed are: 
$\gamma\gamma$, $e^+e^-,$ $\mu^+\mu^-$, $\pi^0\pi^0$, $\pi^+\pi^-$.
Decay width of sgoldstino into photons: 
\begin{equation}
\label{Sgammagamma}
\Gamma(S\to\gamma\gamma)=\left(\frac{\alpha(m_S)
\beta(\alpha(M_{\gamma\gamma}))}{\beta(\alpha(m_S))\alpha(M_{\gamma\gamma})}
\right)^2\frac{m_{S}^3M_{\gamma\gamma}^2}{32\pi F^2}.
\end{equation}
Here the dimensionless multiplicative factor accounts for the 
renormalization group evolution of the photonic operator at different
mass scales.
Lepton channels are: 
\begin{equation}
\label{Sll}
\Gamma(S\to {}l^+l^-)={m_S^3A_l^2\over 16\pi F^2}{m_{l}^2\over m_S^2}\l
1-{4m_{l}^2\over m_S^2}\r^{3/2}.
\end{equation}

Decay into light mesons is provided with gluonic operator at a low energy scale.

\begin{multline}
\Gamma(S\to\pi^0\pi^0)={\alpha^2_s(M_3)\over\beta^2(\alpha_s(M_3))}
{\pi m_S\over
4}{m_S^2M_3^2\over F^2}\\ \left( \!\!1\!
-\!{\beta(\alpha_s(M_3))\over\alpha_s(M_3)}
{9\over 4\pi}{B_0\over
m_S}{m_u+m_d\over m_S}{A_Q\over M_3}\!\right)^{\!\!2}\!
\sqrt{1\!-\!{4m_{\pi^0}^2\over m_S^2}},
\label{StoPiPi}
\end{multline}

\begin{equation}
\Gamma(S\to\pi^0\pi^0)\approx{\alpha^2_s(M_3)\over\beta^2(\alpha_s(M_3))}
{\pi m_S^3M_3^2\over 4F^2}\sqrt{1-{4m_{\pi^0}^2\over m_S^2}},
\end{equation} 
\begin{equation}
\Gamma(S\to\pi^+\pi^-)=2\Gamma(S\to\pi^0\pi^0)\,.
\end{equation}
See Ref.\cite{Astapov:2015otc} for notations for above formulas.
 
\begin{figure}[htb!]
    \centering
\includegraphics[width=0.5\textwidth]{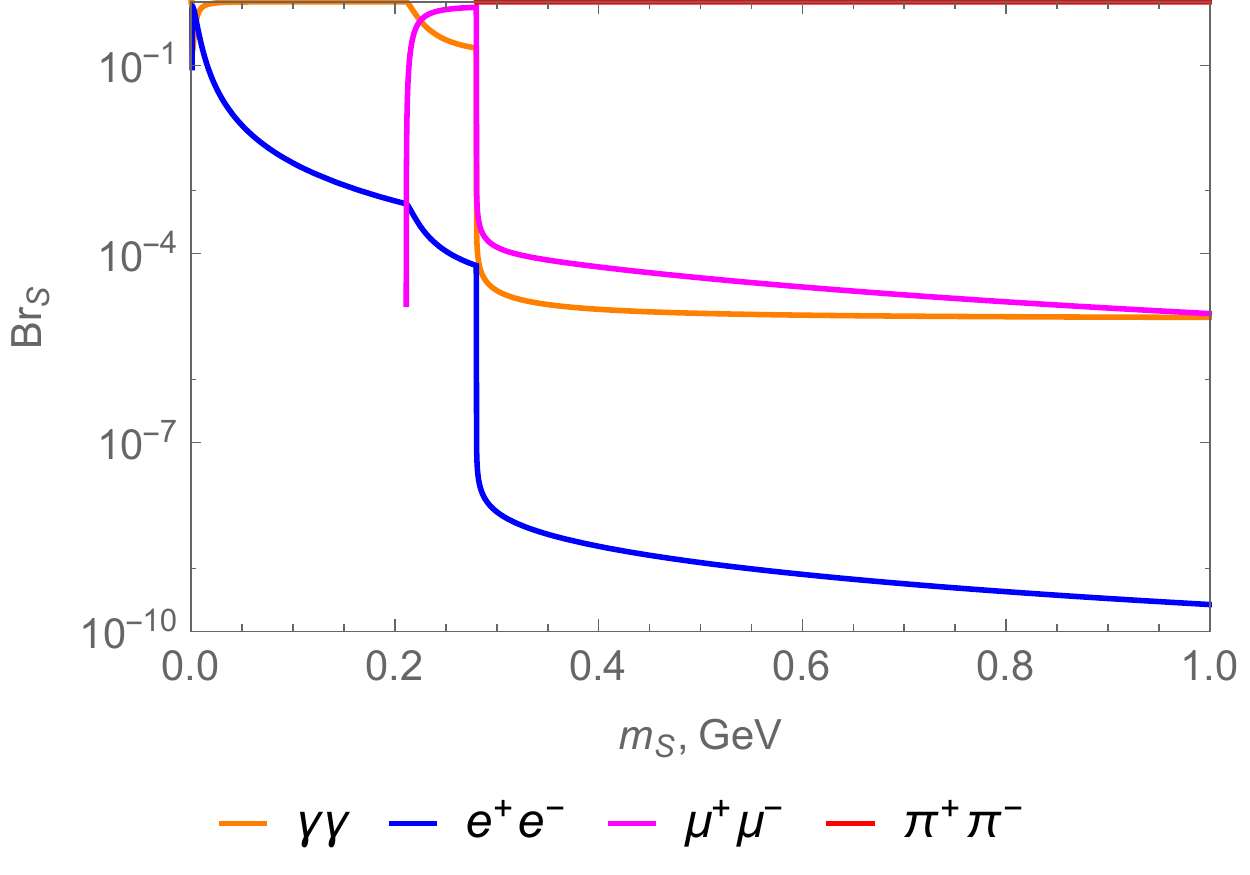}
    \caption{Branching ratios of a scalar sgoldstino.
    \label{SBranching}}
\end{figure}

 Sgoldstino decay branching ratios 
 for the values of MSSM parameters given in Table\,\ref{MSSMpoint}
 are shown in Fig.\,\ref{SBranching}. 
Hadronic channel $\pi\pi$ dominate when it is
kinematically allowed, while
$\gamma\gamma$ and $\mu^+\mu^-$ give small but noticeable
contributions. 
 
\begin{figure}[htb!]
   \centering
\includegraphics[width=0.5\textwidth]{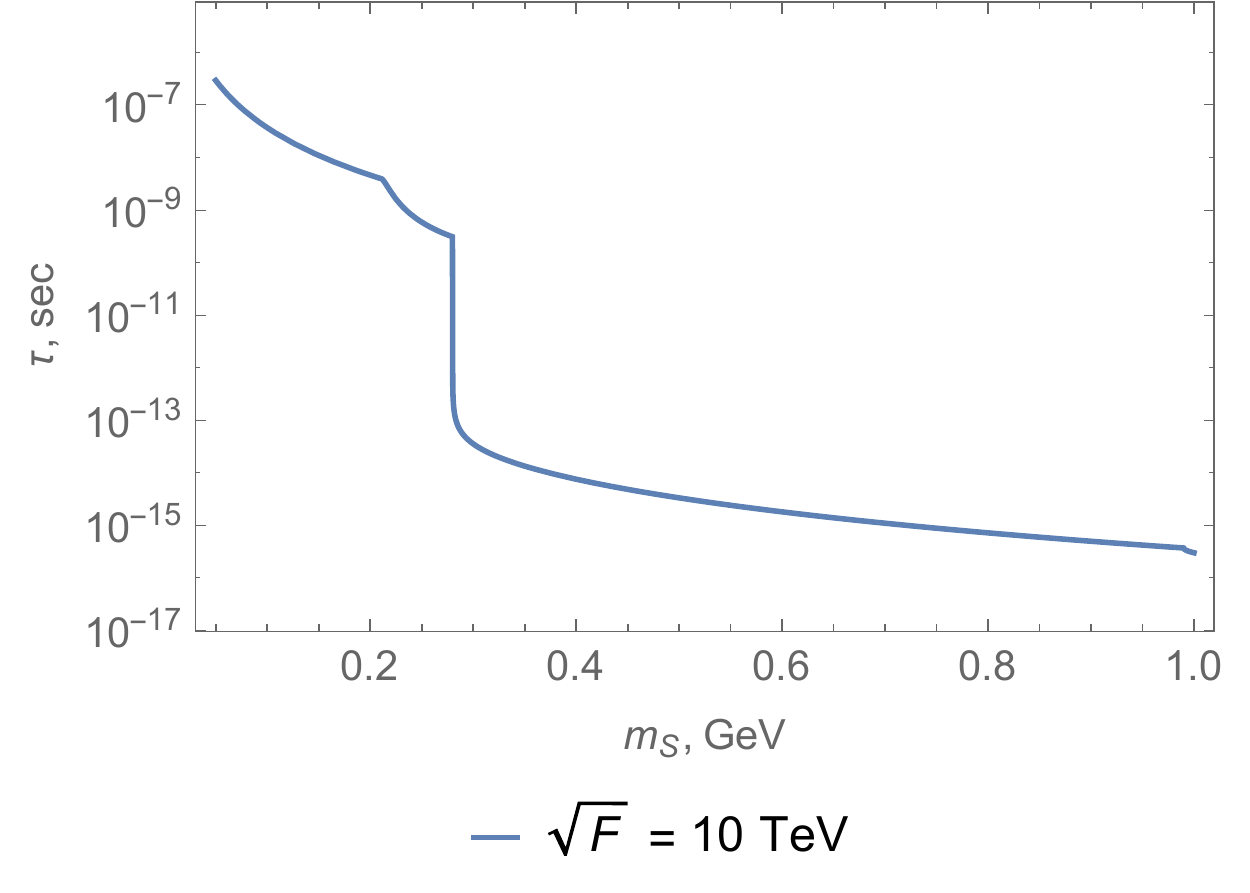}
\caption{Lifetime of a scalar sgoldstino as a function of its mass.
\label{SLifetime}}  
\end{figure}

The sgoldstino lifetime for given $\sqrt{F} = 10$ TeV is 
presented in Fig.\,\ref{SLifetime}. 

\section{Results}
\label{sec:5}
Here we estimate the number of $e^{-}N\to{}S\to{}e^{+}e^{-}$
events inside the fiducial volume of the NA64 experimental setup. 
Experimental setup is designed to search for rare decays with charged particles in the final state.
Detailed setup scheme is outlined in the Ref.\cite{Andreas:2013lya}.
The experiment utilize clean high energy $e^-$ beam with less then $10^{-2}$ level of impurities and
momenta of $100$\,GeV. Electron beam is produced by primary 400 GeV proton beam from SPS hitting the primary beryllium target.  
Electron beam strikes on the electron calorimeter target and produces sgoldstinos directly through the processes described in the previous sections.  
Target calorimeter thickness is $l_{sh}=0.15$\,m. The vacuum vessel length is about 
$l_{det}=15$\,m. It
forms a cylinder along the beam axis with an circle base of
30 cm in diameter. At the back end of the vacuum vessel another electronic calorimeter serves for
count of electromagnetic shower produced by subsequent  sgoldstino decays into charged particles. 

The number of signal events reads
as 

\begin{multline}
N_{\text{signal}}=N_{\text{EOT}}\frac{N_0X_0}{A}\int^{E_0-m_e}_{m_S}dE_S\int^{E_0}_{E_S+m_e}dE_e\\
\times\int^{T}_{0}dt\Bigg[I_e(E_0,E_e,t)\frac{1}{E_e}\frac{d\sigma}{dx_e}\Bigg]w_{det}\text{BR}_{det},
\end{multline}
where the expected number of electons on the target 
is $N_{\text{EOT}}=10^{9}$,  $N_0$ is Avagadro's number, $X_0$ is the unit radiation length of the target material, A is atomic mass number, $E_s$ is sgoldstino energy, $E_0$ and $E_e$ are beam and initial electron energies correspondingly, $x_e=\frac{E_s}{E_e}$  
and $w_{det}$ denotes the probability for the sgoldstino to decay
inside the fiducial volume of the detector,  
\begin{multline}
w_{det}(E_{S(P)}, m_{S(P)}, \sqrt{F})=\exp(-l_{sh}/\gamma{}c\tau_{S(P)})\times\\
\times\left[1-\exp(-l_{det}/\gamma{}c\tau_{S(P)})\right],
\end{multline}
with the sgoldstino gamma factor $\gamma{}=E_{S(P)}/m_{S(P)}$.

Since electron beam with energy $E_0$ becomes degraded as electrons pass trough and interact with its
nucleus. Energy distribution of electrons (see Ref.\cite{Andreas:2012mt}) after passing through material by $t$ radiation length is given by: 
\begin{equation}
I_e(E_0,E_e,t) = \frac{1}{E_0}\frac{\Big[\ln(\frac{E_0}{E_e})\Big]^{bt-1}}{\Gamma(bt)},
\end{equation}
where $\Gamma$ is Gamma function, $b=4/3$, $E_0$ is initial beam energy at $t=0$.

\begin{figure}[!htb]
    \centering
\includegraphics[width=0.45\textwidth]{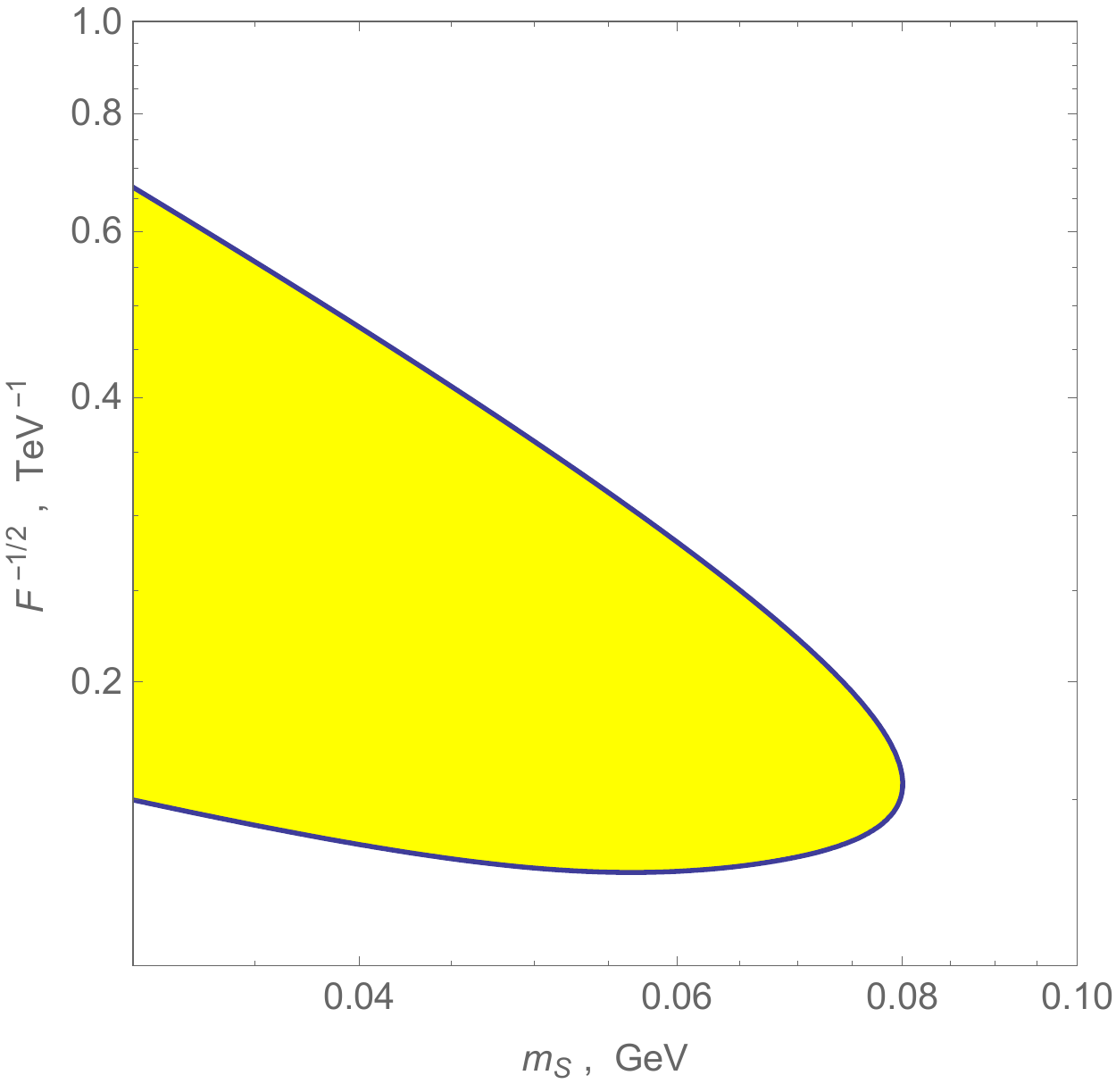}
    \caption{The shaded region  
will be probed at the NA64 experiment. 
    \label{Scalar}}
\end{figure}

In Fig.\,\ref{Scalar} 
we indicate the region in the model parameter space
$(m_{S},1/\sqrt{F})$, where the number of sgoldstino decay events inside
the fiducial volume exceeds 3, $N_{\text{signal}}>3$. That is, if
no events were observed  the region is excluded at
the confidence level of 95\%, in accordance with the Poisson statistics. 
The lower boundary in Fig.\,\ref{Scalar} 
is the region where the couplings are so small that sgoldstinos
escape from the detector without decay. The upper boundary corresponds to case when 
couplings are so large that sgoldstinos decay before the detector.
The scalings of the signal
events imply that models with a higher (as compared to that presented in
Fig.\,\ref{Scalar}) scale of supersymmetry breaking  can be tested if
MSSM parameters $\mu$, $M_{\gamma\gamma}$ are appropriately larger (as compared to
those presented in Table\,\ref{MSSMpoint}).

\section{Conclusions}
\label{conclude}

We have estimated sensitivity of the NA64 experiment to supersymmetric
extensions of the SM where sgoldstinos are light. The experiment
will be able to probe the supersymmetry breaking scale $\sqrt{F}$ up to
$10^4$\,TeV.  We have obtained exclusion regions of the scalar sgoldstino parameter space ($m_S$ vs. $1/\sqrt{F}$).

\paragraph*{Acknowledgments}
We thank D.~Gorbunov, S.~Demidov, S. Gninenko, M. Kirsanov, N. Krasnikov and S. Kulagin for valuable discussions. The work was supported by the RSF Grant No. 14-12-01430.


\end{document}